\documentclass[referee]{raa}           
\usepackage{graphicx}  
\usepackage{amsmath,amssymb}
\usepackage{longtable}
\usepackage{booktabs}
\usepackage{pdflscape} 
\usepackage[pagebackref=true]{hyperref}
\usepackage{natbib}

\begin{document}

  \title{White Dwarfs with Infrared Excess from DESI EDR
}

   \volnopage{Vol.0 (20xx) No.0, 000--000}      
   \setcounter{page}{1}        

   \author{Ke-Yi Wang 
      \inst{1}
   \and Qiong Liu
      \inst{1}
     }

   \institute{College of Physics, Guizhou University, Guiyang 550025, China; {\it qliu1@gzu.edu.cn}
         \vs\no
  \\ {\small Received 20xx month day; accepted 20xx month day}}

\authorrunning{Ke-Yi Wang, Qiong Liu} 
\titlerunning{White Dwarfs with Infrared Excess from DESI EDR} 

 \abstract{Infrared (IR) excess emission around white dwarfs (WDs) is commonly attributed to circumstellar debris disks and/or low-mass companions, providing a unique window into the evolution of planetary systems and binary evolution after the main-sequence stage. Based on a spectroscopically confirmed WD sample from the DESI Early Data Release, we performed a systematic search for IR excess by combining multi-band photometry from SDSS, Pan-STARRS, UKIDSS, 2MASS, and WISE. Using spectral energy distribution (SED) fitting, we initially identified 72 IR-excess candidates and conducted a stringent contamination assessment based on higher-resolution imaging within $6\arcsec$ of each target. After removing sources affected by blending or source confusion, we obtained a final sample of 62 reliable IR excess candidates.
Among them, we identify three candidate WD+M dwarf binaries (two new systems), five candidate WD+brown dwarf (BD) binaries (all new), 38 candidate WD+dust disks (28 new), and 16 ambiguous systems that could be either WD+BD or WD+dust (15 new). Compared with previous samples, our catalog extends the parameter space of known dusty WDs toward older cooling ages. Due to the limited spatial resolution of WISE, follow-up high-resolution imaging and/or infrared spectroscopy is required to confirm the physical nature of all candidate systems and to further expand the parameter space of dust disks in terms of cooling age and other properties.}

\maketitle

\section{Introduction}          
\label{sect:intro}
White dwarfs (WDs) represent the final evolutionary stage of the vast majority of stars and offer a valuable window into the long-term evolution of planetary systems and binary companions beyond the main sequence \citep{1997ApJ...489..772I}. Owing to their high effective temperatures, WD emission peaks primarily at optical wavelengths, and their intrinsic infrared (IR) flux is typically weak. Any additional IR-emitting component in the vicinity of a WD—such as a circumstellar dust disk or a low-mass companion—can therefore produce a detectable IR excess that is readily captured by modern IR facilities. 

IR excesses around WDs are generally attributed to two main physical origins: (i) debris disks formed from tidally disrupted planetary remnants, and (ii) binary systems containing a cool, low-mass companion (e.g., an M dwarf or a brown dwarf, BD) whose emission dominates at longer wavelengths. Systematic investigations of IR-excess WDs thus probe both the dynamical fate of planetary material and the survival of low-mass companions through post-main-sequence evolution, providing insights into stellar late-time evolution and its circumstellar environment.

Binary systems composed of a WD and a low-mass companion serve as important laboratories for studying binary evolution and the common-envelope (CE) phase \citep{1993A&A...267..397D}. In WD+M dwarf systems, evolutionary outcomes depend strongly on the initial orbital separation\citep{1993A&A...267..397D,2004A&A...419.1057W,2025A&A...695A.161S}. Systems with wide initial separations ($\gtrsim 10$~AU) typically evolve with little interaction, allowing the primary to form a WD while the orbit expands due to mass loss \citep{2010ApJS..190..275F}. In closer configurations, the primary may engulf the companion during its red-giant phase, leading to a CE episode \citep{1993PASP..105.1373I}. Envelope ejection subsequently produces compact post-common-envelope binaries (PCEBs) \citep{2012MNRAS.423..320R, 2014A&A...568A..68Z}, which are key systems for constraining CE efficiency and may evolve into cataclysmic variables, double-degenerate binaries, Type~Ia supernova progenitors, or low-mass X-ray binaries \citep{2004MNRAS.350.1301H, 2012NewAR..56..122W, 2013MNRAS.429..256P, 2021ApJS..257...65S, 2010A&A...520A..86Z, 2022MNRAS.512.1843H}.

BDs, with masses of $\sim 13$--$80\,M_{\rm Jup}$, are substellar objects incapable of sustaining stable hydrogen fusion \citep{2000ARA&A..38..337C, 2011ApJ...736...47B}. Close BD companions are already rare around main-sequence stars and appear even less frequently around WDs. As in WD+M systems, orbital separation plays a central role in their evolution: wide systems may remain largely unaffected by post-main-sequence evolution, whereas close systems likely underwent a CE phase during the red-giant expansion of the WD progenitor \citep{maxted2006nature, 2017MNRAS.471..948R}. Because BDs radiate most strongly in the IR, WD+BD systems often exhibit distinctive infrared SEDs, and their spectra may show molecular absorption features such as CH$_4$ and H$_2$O \citep{1996ApJ...467L.101G, 1998ApJ...502..932O}. Statistical studies of WD+M and WD+BD binaries therefore provide valuable constraints on CE physics and on the survival of low-mass companions through stellar evolution.

In contrast, WD+debris-disk systems are linked to the dynamical evolution of planetary remnants. Minor bodies can be perturbed onto highly eccentric orbits by surviving planets \citep{1990ApJ...357..216G, 2003ApJ...584L..91J, 2012ApJ...747..148D}, eventually crossing the WD Roche limit and undergoing tidal disruption \citep{1999Icar..142..525D}. The resulting debris forms a circumstellar disk. Both theoretical models and observations indicate that some planets and rocky bodies can survive the red-giant phase and remain gravitationally bound after WD formation \citep{2007ApJ...661.1192V, 2009ApJ...705L..81V, 2010MNRAS.408..631N, 2012ApJ...761..121M}. However, the presence of dust in close proximity to WDs implies ongoing or recurrent delivery mechanisms, including repeated tidal disruptions, long-term dynamical perturbations, or instabilities in multi-planet systems \citep{2013MNRAS.431.1686V, 2015Natur.526..546V, 2015MNRAS.454...53B, 2016ApJ...816L..22X, 2018MNRAS.481.2601F, 2017MNRAS.469.2750S, 2023MNRAS.524.6181O}. Dust grains gradually spiral inward under Poynting–Robertson drag \citep{2011ApJ...732L...3R}, sublimate near the WD, and may accrete onto the stellar surface \citep{2012ApJ...760..123R, 2014MNRAS.442L..71V}. Since heavy elements sink rapidly out of WD atmospheres, the detection of metal pollution indicates ongoing or recent accretion \citep{2009A&A...498..517K}. Approximately 40\% of WDs exhibit atmospheric metal contamination, supporting the widespread presence of remnant planetary material \citep{2003ApJ...596..477Z, 2010ApJ...722..725Z, 2014A&A...566A..34K}.

Large-scale photometric surveys have enabled systematic searches for IR excess around WDs. Early studies based on SDSS catalogs established the observational foundation for identifying dusty disks and cool companions \citep{2011ApJS..197...38D, 2011MNRAS.417.1210G}. The advent of Gaia, with precise parallaxes and improved photometric homogeneity, significantly enhanced the reliability of such searches. For instance, \cite{2020ApJ...902..127X} combined Gaia and unWISE/\textit{WISE} photometry to identify 188 IR-excess candidates among bright WDs, substantially expanding the known parameter space. Similar approaches applied to spectroscopic surveys such as LAMOST DR5 have extended searches toward fainter and more distant systems \citep{2023ApJ...944...23W}.

The Dark Energy Spectroscopic Instrument (DESI) provides a new opportunity to advance this field. Its large aperture and high-multiplex capability yield high-quality spectra for vast numbers of targets, including a substantial sample of spectroscopically confirmed WDs in the Early Data Release (EDR). The depth of DESI allows access to fainter and more distant WDs, extending searches into regimes of lower luminosity and potentially older cooling ages. Spectroscopic confirmation further improves sample purity compared to purely photometric selections, offering a secure basis for subsequent SED analysis. When combined with wide-area optical and IR surveys, DESI enables the construction of a homogeneous and statistically robust catalog of IR-excess candidates and defines well-motivated targets for follow-up observations.

Motivated by these considerations, we conduct a systematic search for IR excess among 2706 spectroscopically confirmed WDs in the DESI EDR by cross-matching with multi-wavelength optical and IR surveys. We assemble a homogeneous photometric dataset spanning optical to IR wavelengths, identify IR-excess signatures through SED fitting, and assess candidate reliability via imaging-based contamination checks. Two-component modeling is then employed to investigate the physical origin of the excess emission. This strategy yields a robust catalog of IR-excess WD candidates and establishes a prioritized sample for future IR spectroscopy and high-resolution follow-up observations.

This paper is organized as follows. Section~\ref{sect:Obs} describes the DESI WD sample and the cross-matching procedures. Section~\ref{sect:SED} presents the SED-fitting methodology and selection criteria. Section~\ref{sect:image} details the imaging inspection and classification scheme. Section~\ref{sect:DISCUSSION} reports the results and discusses the physical and statistical properties of different subsamples. Section~\ref{sect:conclusion} summarizes our conclusions and outlines prospects for future work.

\section{Photometric Data Collection}
\label{sect:Obs}
Our analysis is based on a catalog of 2706 spectroscopically confirmed WDs from the DESI Early Data Release (EDR; \cite{2024MNRAS.535..254M}). This catalog excludes known binary and interacting systems—including WD+MS pairs, double degenerates (WD+WD), and cataclysmic variables—so that the resulting sample is dominated by single WDs. The spectra were obtained with the DESI instrument on the Mayall 4\,m telescope at Kitt Peak National Observatory between 2020 December 14 and 2021 June 10, covering 3600--9824\,\AA\ with a median spectral resolution (FWHM) of $\simeq 1.8$\,\AA.

Starting from these 2706 spectroscopically confirmed WDs, we cross-match the sample with the WD candidate catalog of \cite{2021MNRAS.508.3877G}. Adopting their reliability definition, we apply a selection threshold of $P_{\rm wd} > 0.75$, a commonly used criterion for robust WD identification. Although the DESI sample is already spectroscopically confirmed, we retain this cut as an additional quality-control step to exclude a small number of potentially contaminated objects. This step removes 36 objects, leaving a parent sample of 2670 WDs for the subsequent search for IR excess.

To investigate potential IR excesses, we compile multi-wavelength photometry by cross-matching the parent sample with major optical and infrared surveys via VizieR\footnote{\url{https://vizier.cds.unistra.fr}} using a matching radius of $3\arcsec$. In the optical regime, we cross-match simultaneously with both Pan-STARRS DR1 and SDSS DR12. This approach ensures that each WD in the parent sample is covered by at least one set of optical measurements, while the majority of sources are associated with photometry from both surveys.

In the IR, we cross-matched our sample with the ALLWISE catalog by propagating the WD coordinates to the ALLWISE epoch (2010.0), obtaining 431 matches. We then cross-matched the same sample with the CatWISE2020 catalog by propagating the WD coordinates to the CatWISE2020 epoch (2015.4), obtaining 1070 matches. Combining the two catalogs yields a total of 1087 WDs with WISE counterparts. Among these, 17 sources have CatWISE2020 counterparts but fall outside our adopted 3\arcsec matching radius; for these objects, we include the WISE association via their ALLWISE counterparts. For the W1 and W2 bands, we preferentially adopt photometry from CatWISE2020, which generally provides improved precision compared to ALLWISE. For the longer-wavelength W3 and W4 bands, which are not provided by CatWISE2020, we adopt the corresponding measurements from ALLWISE.
For the near-IR (NIR) data, we first cross-matched the 1087 WISE-matched WDs with UKIDSS DR9 by propagating the WD coordinates to the representative UKIDSS epoch (2016.0), obtaining 224 counterparts. We then cross-matched the same sample with 2MASS by propagating the WD coordinates to the 2MASS reference epoch (J2000.0), obtaining 112 additional detections. Twenty sources have NIR measurements in both UKIDSS and 2MASS; for these, we preferentially adopt the UKIDSS photometry.

After requiring the presence of optical, IR photometry, we obtain a final working sample of 316 WDs, which forms the basis for SED construction and IR-excess identification.

\section{VOSA SED Fitting}
\label{sect:SED}
We used VOSA\footnote{\url{http://svo.cab.inta-csic.es/theory/vosa/}} to perform SED fitting for the 316 spectroscopically confirmed WDs with complete optical, IR photometry. For each source, we compiled multi-band photometry together with the Gaia EDR3 distance and line-of-sight extinction. The SEDs were corrected for interstellar extinction using the law of \cite{1999PASP..111...63F}, including the infrared modifications from \cite{2005ApJ...619..931I}. 

The photometric data were fitted with the Koester WD atmosphere models implemented in VOSA \citep{2010MmSAI..81..921K}. For the majority of sources, the optical and IR measurements are well reproduced by the WD model, while deviations typically emerge at longer wavelengths. Only a small number of objects show excess emission already in the NIR; in most cases, the first significant departure from the pure WD model appears in the WISE W1 band, indicating that the excess signal is generally confined to the IR regime.

During the excess-identification process, we noted that some sources exhibit marginal flux deviations that could arise from underestimated photometric uncertainties or residual systematics. To minimize spurious detections, we adopt a conservative threshold by combining the reported photometric uncertainties with additional model and calibration terms. A data point is considered to show significant IR excess if
\begin{equation}
F_{\mathrm{obs}} - F_{\mathrm{mod}} > 3\sigma_{\mathrm{tot}} ,
\label{eq:basic_threshold}
\end{equation}
where the total uncertainty is defined as
\begin{equation}
\sigma_{\mathrm{tot}} = \sqrt{\sigma_{\mathrm{obs}}^{2} + \sigma_{\mathrm{mod}}^{2} + \sigma_{\mathrm{cal}}^{2}} .
\label{eq:total_uncertainty}
\end{equation}

Here, $\sigma_{\mathrm{obs}}$ denotes the catalog photometric uncertainty, $\sigma_{\mathrm{mod}}$ represents the uncertainty in the WD model prediction (taken as 5\% of the observed flux following \cite{2020ApJ...902..127X}), and $\sigma_{\mathrm{cal}}$ is the absolute calibration uncertainty of the corresponding survey. We adopt calibration uncertainties of 2.4\% for WISE W1 and 2.8\% for WISE W2 \citep{2011ApJ...735..112J}, and 1.685\% for 2MASS $K_s$ / UKIDSS $K$ \citep{2003AJ....126.1090C}. These correspond to
$\sigma_{\mathrm{cal},W1} = 0.024\,F_{\mathrm{obs},W1}$,
$\sigma_{\mathrm{cal},W2} = 0.028\,F_{\mathrm{obs},W2}$, and 
$\sigma_{\mathrm{cal},K} = 0.01685\,F_{\mathrm{obs},K}$.

Applying this criterion to the SED fitting results yields 72 initial WD candidates with significant IR excess.

\section{Results of Infrared Excess Characterization}
\label{sect:image}

In this section, we characterise the IR excess detected in the WD candidates identified from the SED fitting analysis. The analysis focuses on evaluating the reliability of the detected excess and on investigating its physical origin. To this end, we adopt a two-step approach in which multi-wavelength imaging is used to identify and exclude spurious excesses caused by source confusion or background contamination, followed by two-component SED fitting of the remaining reliable systems.

\subsection{Image-based Contamination Assessment}

\begin{figure*}
    \centering
    \begin{minipage}{0.32\textwidth}
        \centering
        \textbf{WISE W1} \\  
        \includegraphics[width=\linewidth]{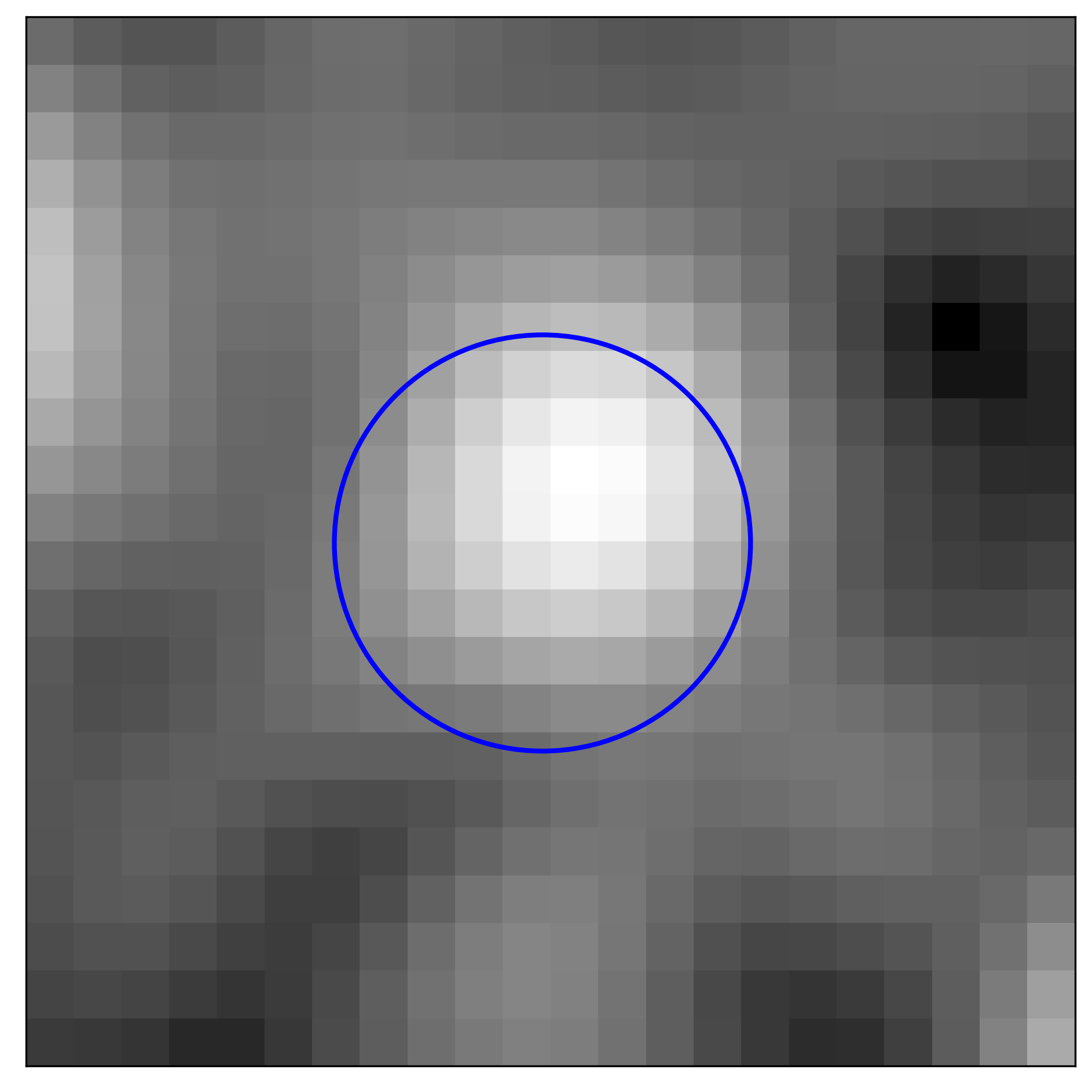} 
    \end{minipage}
    \begin{minipage}{0.32\textwidth}
        \centering
        \textbf{UKIDSS K} \\ 
        \includegraphics[width=\linewidth]{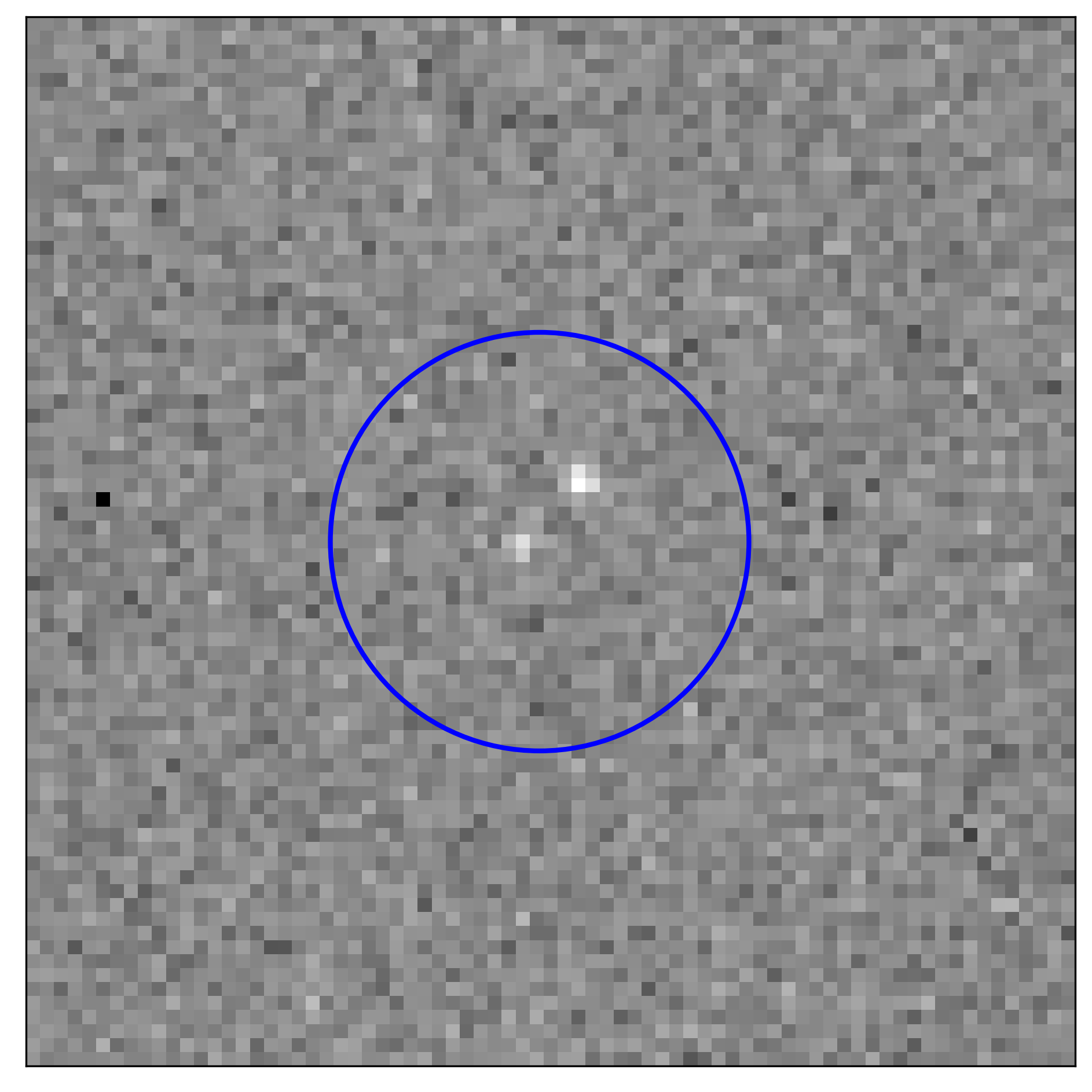} 
    \end{minipage}
    \begin{minipage}{0.32\textwidth}
        \centering
        \textbf{Pan-STARRS z} \\  
        \includegraphics[width=\linewidth]{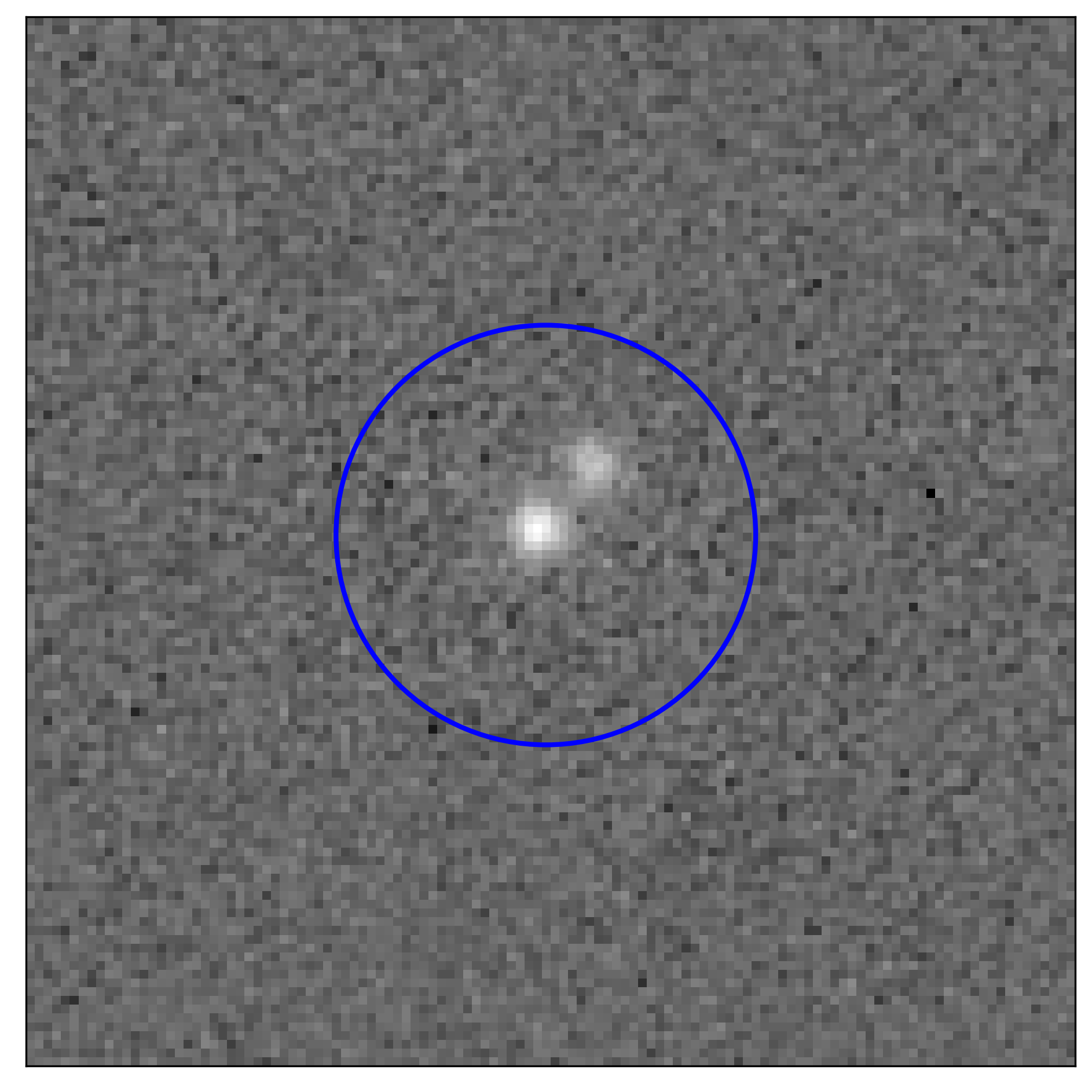} 
    \end{minipage}
    \caption{Images of Gaia DR3 1468336588498467328. The left panel shows the \textbf{WISE W1} (3.4\,\(\mu\)m) image, the middle panel presents the \textbf{UKIDSS \(K\)-band} (2.2\,\(\mu\)m) image, and the right panel displays the \textbf{Pan-STARRS \(z\)-band} (optical) image. Each cutout covers \(30\arcsec \times 30\arcsec\). The blue circle, centered on the DESI coordinate of the WD, indicates a radius of \(6\arcsec\). Both the Pan-STARRS and UKIDSS images show additional sources within this radius, implying that the photometry of the WD may be affected by blending and that the inferred IR excess could be contaminated by nearby objects.}
    \label{fig:confused_wise}
\end{figure*}

The intrinsic faintness of WDs, together with the relatively large point-spread function (PSF) of the \textit{WISE} instrument, renders IR photometry particularly vulnerable to contamination from unresolved nearby sources. Such contamination can produce apparent weak IR excesses and, if not properly accounted for, lead to false-positive detections. A careful inspection of the surrounding field is therefore essential for constructing a robust sample of IR-excess WDs.

For each of the 72 initial IR-excess candidates, we examined the region within a $6\arcsec$ radius centred on the WD position. This radius is comparable to the typical \textit{WISE} PSF sizes in the W1 ($6.1''$) and W2 ($6.4''$) bands and thus defines the spatial scale over which blending is most likely to occur. We inspected higher-resolution optical images from the SDSS and Pan-STARRS $z$ bands, as well as NIR images from the 2MASS $K_s$ and UKIDSS $K$ bands, and compared them directly with the corresponding \textit{WISE} W1 images.

A candidate was classified as contaminated if one or more neighbouring sources were clearly resolved within the $6\arcsec$ radius in the higher-resolution images, indicating that the \textit{WISE} photometry may be affected by blended flux. Figure~\ref{fig:confused_wise} presents a representative case in which the target appears as a single source in the \textit{WISE} image but is resolved into multiple components in the UKIDSS and Pan-STARRS images.

Applying this criterion, we identified 10 candidates affected by source confusion and excluded them from further analysis. The final sample therefore consists of 62 reliable IR-excess WD candidates. We note that this image-based vetting is intentionally conservative, aimed at removing clear false positives rather than ensuring a completely uncontaminated sample, given the intrinsic spatial-resolution limitations of \textit{WISE}.

\subsection{Two-component SED Fitting and Classification}

To investigate the physical origin of the IR excess in the 62 vetted candidates, we performed two-component SED fitting within the VOSA framework. In addition to the WD photospheric emission described by the Koester atmosphere models \citep{2010MmSAI..81..921K}, an extra component was introduced to account for the observed IR emission. This excess component was modelled in two complementary ways: a single-temperature blackbody, representing thermal emission from circumstellar dust, and the BT-Settl atmosphere models, representing low-mass stellar or substellar companions.

The excess component was first characterised using the blackbody model, from which a characteristic blackbody temperature, $T_{\rm BB}$, was derived. Based on the expected dust sublimation temperatures and the typical effective temperatures of low-mass companions, we adopted the following classification scheme. Sources with $T_{\rm BB} > 2100\,{\rm K}$ are likely dominated by companion emission, while those with $T_{\rm BB} < 1200\,{\rm K}$ are attributed to circumstellar dust disks. Objects with intermediate temperatures ($1200\,{\rm K} \leq T_{\rm BB} \leq 2100\,{\rm K}$) are classified as ambiguous, as their IR excess can plausibly originate from either warm dust or a cool substellar companion.

For systems classified as companion-dominated or ambiguous on the basis of the blackbody fits, we performed refined two-component fitting using the BT-Settl (CIFIST) models \citep{2011ASPC..448...91A} to estimate the companion effective temperature, $T_{\rm BT}$. An empirical boundary of $T_{\rm BT} = 2500\,{\rm K}$ was adopted to distinguish M-dwarf companions from BDs. Systems with $T_{\rm BT} \geq 2500\,{\rm K}$ were classified as WD+M dwarf candidates, whereas those with $T_{\rm BT} < 2500\,{\rm K}$ were classified as WD+BD candidates.

Applying this two-stage classification procedure, the final sample of 62 IR-excess WDs comprises three WD+M dwarf binary candidates, five WD+BD binary candidates, 38 WD+dust disk candidates, and 16 systems whose nature cannot be uniquely determined with the current photometric constraints. These ambiguous systems are retained as promising targets for future high-resolution imaging and IR spectroscopic follow-up, which will be essential for breaking the companion--disk degeneracy.

We emphasize that the ``ambiguous'' category is defined solely by the blackbody temperature range $1200 \leq T_{\rm BB} \leq 2100$~K. The subsequent BT-Settl fitting is used only to explore a possible companion interpretation, and does not alter the original ambiguous classification.

\section{DISCUSSION}\label{sect:DISCUSSION}
The fraction of systems in our final sample whose IR excess is dominated by a companion is relatively small. This is consistent with both the DESI target selection and the construction of our parent catalog. DESI spectroscopy is optimized for predominantly faint sources, and most objects in our sample have Gaia $G$-band magnitudes of $G \gtrsim 18.5$. At these magnitudes, the optical contribution from low-mass companions (e.g., M dwarfs or BDs) is intrinsically weak and further suppressed by photometric uncertainties and detection limits, making such systems difficult to recognize in the optical. In addition, our starting catalog of 2706 spectroscopically confirmed WDs from the DESI EDR was compiled to exclude previously identified binary and interacting systems (e.g., WD+MS, WD+WD, and cataclysmic variables). Taken together, these factors naturally lead to a low incidence of companion-dominated candidates in the final IR-excess sample.

The robustness of an apparent IR excess also depends on wavelength. In the WISE bands, the modest angular resolution and large point-spread function of WISE increase susceptibility to source confusion, contamination by nearby stars or background galaxies, and systematic photometric biases, particularly in crowded fields. IR excesses identified solely from WISE photometry should therefore be regarded as candidate signals, with their reality and physical origin (dust disk versus cool companion) requiring confirmation through higher-resolution imaging, deeper IR photometry, and/or IR spectroscopy. By contrast, objects exhibiting a significant NIR excess (e.g., in UKIDSS/2MASS) generally provide stronger evidence for a genuine companion contribution, given the typically higher spatial resolution in the NIR and the more direct flux contribution from low-mass companions in these bands.

\begin{figure}
    \centering
    \includegraphics[width=0.75\linewidth]{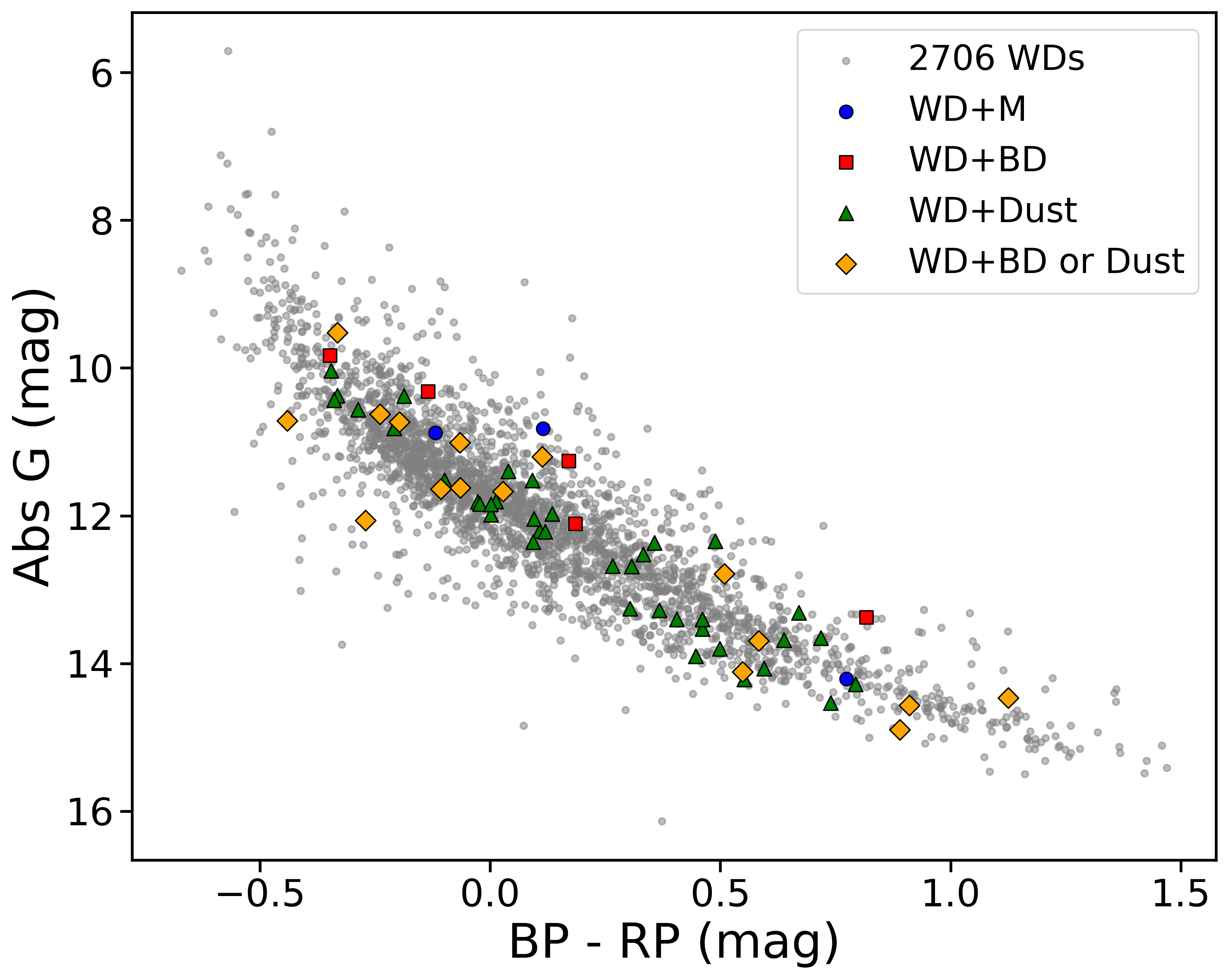}
    \caption{Gaia Hertzsprung–Russell (HR) diagram for the 2706 spectroscopically confirmed WDs from the DESI EDR sample. Grey points show the distribution of the full sample in absolute $G$ magnitude ($M_G$) versus $(BP-RP)$ color. The colored symbols highlight the 62 IR excess candidates classified from two-component SED fitting: WD+M (blue circles), WD+BD (red squares), WD+Dust (green triangles), and ambiguous systems that may be either WD+BD or WD+Dust (orange diamonds).}
    \label{fig:HR}
\end{figure}

To place our candidates in a broader context, we compare their distribution with that of the parent WD sample on the Gaia Hertzsprung--Russell (HR) diagram. Figure~\ref{fig:HR} shows the locations of both the initial WD sample and the final IR-excess candidates in the color--magnitude plane, offering a clear view of how the candidates populate parameter space relative to the parent population.

In the following subsections, we present a detailed classification and analysis of the four classes of IR-excess systems, emphasizing their physical interpretation and statistical properties.

\subsection{WD+M Dwarf Binary Candidates}
\begin{figure}
    \centering
    \includegraphics[width=0.75\linewidth]{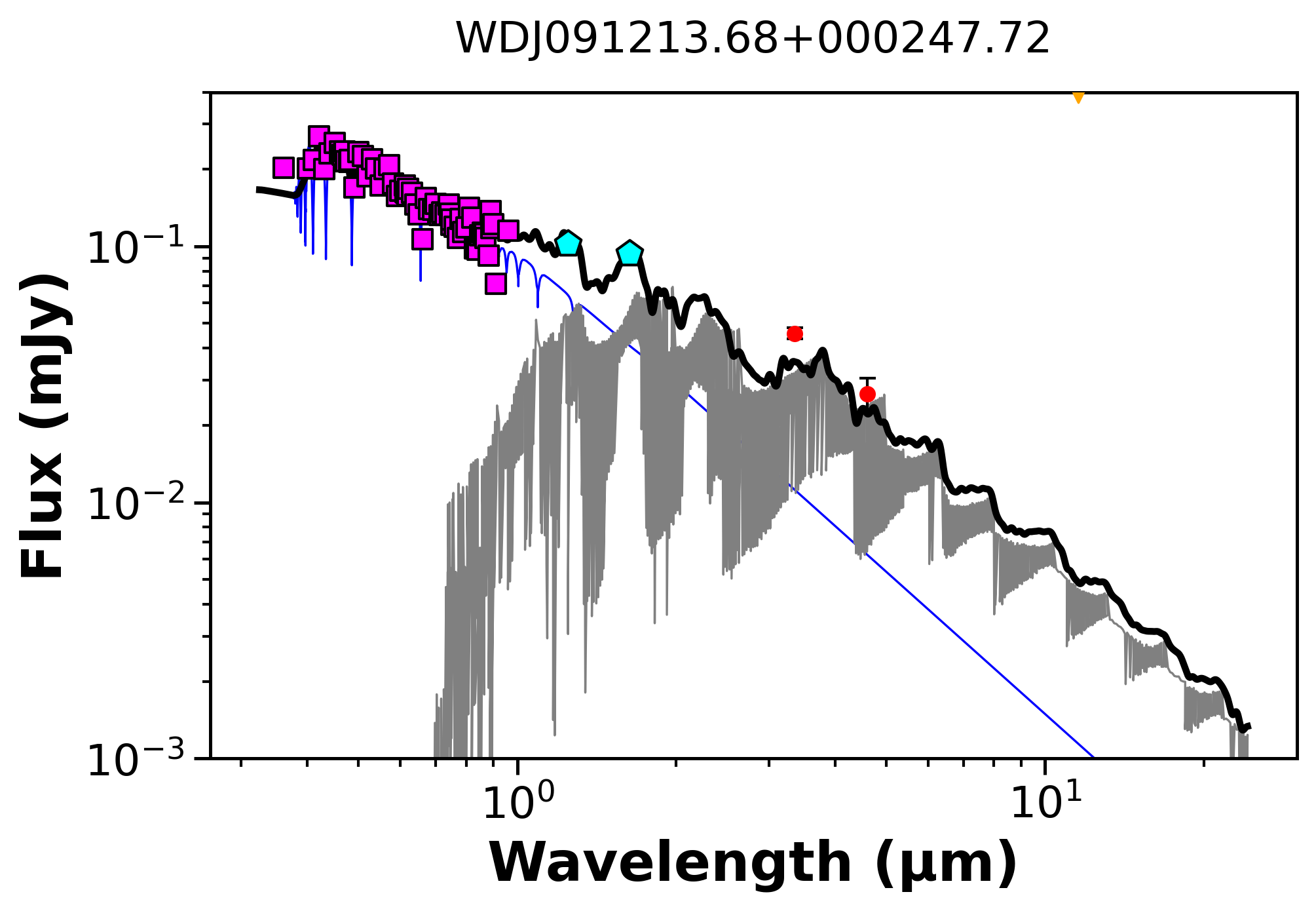}
    \caption{SED fitting result for the WD+M dwarf candidate (WDJ091213.68+000247.72).
The best-fit model (black line) combines a WD (Koester;  blue line) and an M-type companion (BT-Settl; gray line), yielding reduced $\chi^2 = 5.2$.
Photometry includes: SDSS or  Pan-STARRS (magenta squares), UKIDSS or 2MASS (cyan pentagons), WISE (red circles).}
    \label{fig:WD+M}
\end{figure}

We identify a total of three WD+M dwarf binary candidates, which are listed in the upper part of Table~\ref{tab:two_samples}. As illustrated in Figure~\ref{fig:WD+M}, the IR excess in these systems typically becomes evident from the NIR bands onward, in agreement with theoretical expectations. Because the SED of an M dwarf peaks in the NIR, the companion contributes substantially at these wavelengths, resulting in observed NIR fluxes that lie well above those predicted by a pure WD atmosphere model. In contrast, the optical photometry of these systems is generally well reproduced by the WD model alone, indicating that the companion contributes only weakly at optical wavelengths and that the IR excess is dominated by the M-dwarf component. This characteristic SED morphology provides a robust diagnostic for distinguishing WD+M binaries from dust-disk systems. Moreover, the excess emission in WD+M systems typically appears as a smooth continuum enhancement across multiple NIR bands, rather than the more localised ``warm-dust''-like rise commonly observed in disk-hosting WDs, further supporting the reliability of our classification.

As an additional consistency check, we make use of the Gaia low-resolution BP/RP spectrophotometric data and convert them into the J-PAS(Javalambre Physics of the Accelerating Universe Astrophysical Survey; \cite{2014arXiv1403.5237B}) narrow-band photometric system using the \texttt{GaiaXPy} package \citep{2024zndo..11617977R}. The synthetic J-PAS photometry is generated following the methodology adopted in previous studies (e.g. \citet{2025A&A...699A.153R}). The resulting photometry is broadly consistent with the interpretation inferred from the broadband SED fitting, showing no significant excess at optical wavelengths and a clear deviation only toward longer wavelengths. Such behaviour is fully consistent with expectations for WD+M binaries, in which the companion contribution becomes appreciable primarily in the NIR. We emphasize that the Gaia-to-J-PAS transformation is used here solely as a qualitative verification and does not play a role in the formal classification of the systems.

Among the three WD+M binary candidates identified in this work, only one object (WDJ083254.36+313904.84) has been previously reported by \cite{2011ApJS..197...38D}. Owing to the lack of 2MASS measurements, that study relied mainly on SDSS and \textit{WISE} photometry and suggested a possible L0-type companion. In our analysis, the inclusion of higher-precision UKIDSS NIR data provides tighter constraints on the excess component. The blackbody fit yields $T_{\rm BB} > 2100\,\mathrm{K}$, favouring a companion-dominated interpretation, while the BT-Settl fit gives $T_{\rm BT} \gtrsim 2500\,\mathrm{K}$, indicating that an M-dwarf companion is more likely than an L-type object. Nevertheless, broadband SED fitting alone cannot uniquely determine the companion spectral subtype owing to residual model degeneracies and photometric uncertainties; near-infrared spectroscopy will therefore be required for a definitive classification.

The occurrence rate of WD+M dwarf binary candidates in our sample is $3/316 \simeq 0.94\%$. This relatively low fraction is likely due not only to the faint-magnitude nature of the DESI WD sample, which limits the detectability of low-mass companions, but also to the fact that the initial DESI WD catalog was constructed to exclude previously identified WD+MS binaries. Consequently, the lower incidence compared with the typical WD+M binary fraction of $\sim 10\%$ reported in the literature is most likely driven by selection effects rather than reflecting a genuinely lower intrinsic binary fraction.

\begin{figure}
    \centering
    \includegraphics[width=0.75\linewidth]{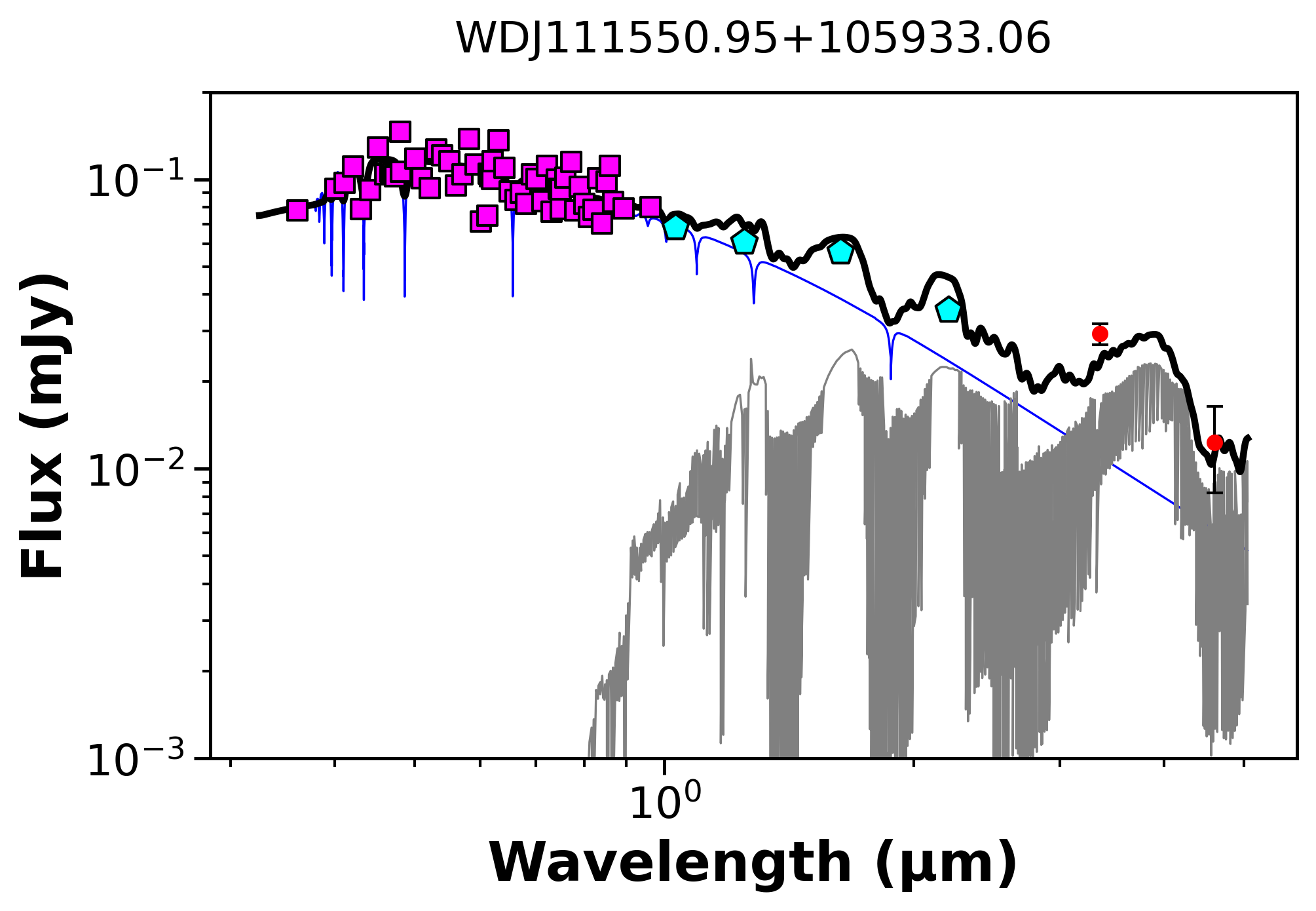}
    \caption{SED fitting result for the WD+BD dwarf candidate (WDJ111550.95+105933.06).
The best-fit model (black line) combines a WD (Koester;  blue line) and an BD-type companion (BT-Settl; gray line), yielding reduced $\chi^2 = 2.0$. Photometry includes: SDSS or  Pan-STARRS (magenta squares), UKIDSS or 2MASS (cyan pentagons), WISE (red circles).}
    \label{fig:WD+BD}
\end{figure}

\begin{figure*}
    \centering
    \begin{minipage}{0.496\textwidth}
        \centering
        \includegraphics[width=\linewidth]{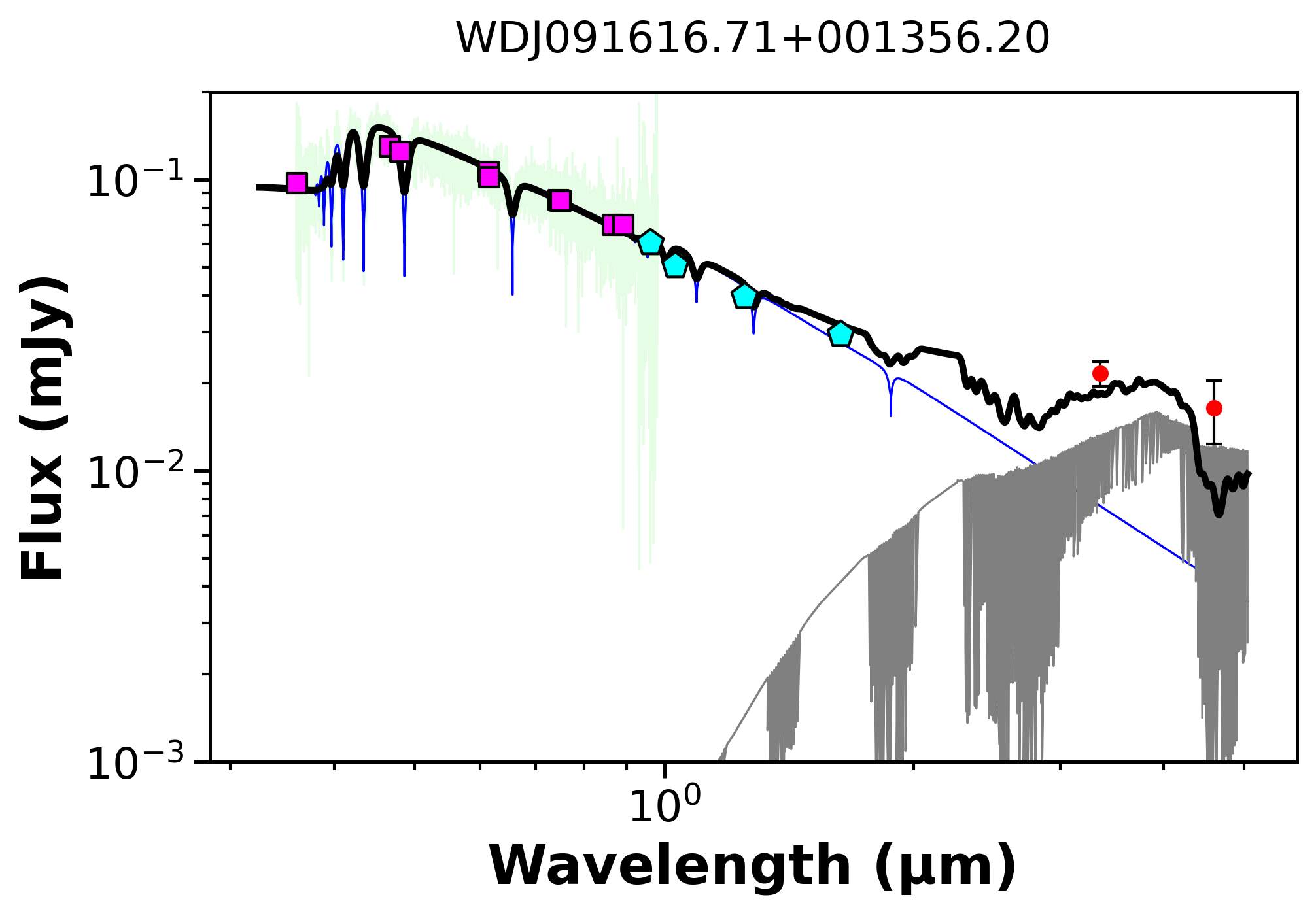} 
    \end{minipage}
    \begin{minipage}{0.496\textwidth}
        \centering
        \includegraphics[width=\linewidth]{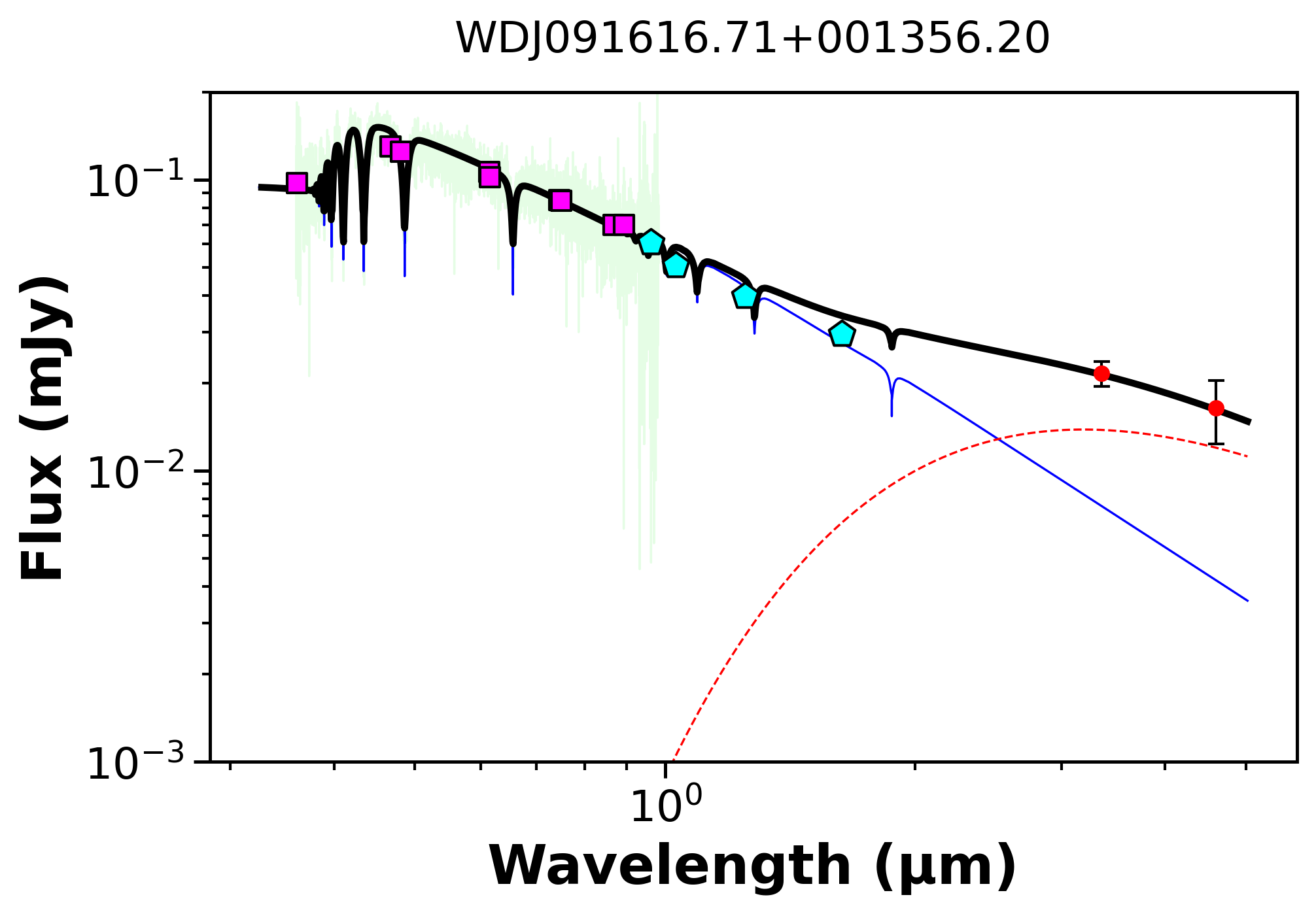} 
    \end{minipage}
    \caption{SED fitting for the source WDJ091616.71+001356.20, illustrating two alternative interpretations of the IR excess. The left panel shows a WD+companion model (Koester + BT-Settl; $T_{\mathrm{BT}} \approx 1500$~K, reduced $\chi^2 = 3.9$), while the right panel shows a WD+dust disk model (Koester + blackbody; $T_{\mathrm{BB}} \approx 1400$~K, reduced $\chi^2 = 4.3$). Photometric data are taken from SDSS (magenta squares), UKIDSS (cyan pentagons), and \textit{WISE} (red circles). The light-green spectrum represents the DESI optical spectrum of the source, which is well reproduced by the WD model in both cases. Both models provide comparably good fits to the available data, leaving the physical origin of the IR excess—circumstellar dust or a low-mass companion—ambiguous.
}
\label{fig:BD_or_dust}
\end{figure*}

\begin{table*}
\centering
\scriptsize
\caption{WD+M dwarf/BD Binary Candidates}
\label{tab:two_samples}
\begin{tabular}{lcccccccccc}
\hline
Name & R.A. & Decl. & $G$ & SpT & $T_{\rm eff}$ & $\log g$ & Dis & $\log(L/L_{\odot})$ & $T_{BT}$ & $\chi^2_{BT}$ \\
 & (deg) & (deg) & (mag) &  & (K) & (cgs) & (pc) &  & (K) &  \\
\hline
WDJ083254.36+313904.84$^a$ & 128.227008 & 31.650450 & 17.8 & DAZ & 5774 & 8.00 & 51.20 &-4.74& 2700 & 32 \\
WDJ084517.78+100059.91 & 131.324097 & 10.016591 & 19.6 & DA & 10228 & 7.11 & 505.35 & -2.74& 2600 & 4.1 \\
WDJ091213.68+000247.72 & 138.056818 & 0.046678 & 18.2 & DA & 14314 & 7.63 & 297.54 &-3.43& 2700 & 5.2 \\

\hline\hline   

WDJ022858.80-005031.27 & 37.245006 & -0.841986 & 19.7 & DA & 16941 & 7.44 & 776.82 &-3.09 & 2150 & 2.0 \\
WDJ111550.95+105933.06 & 168.962269 & 10.992358 & 18.8 & DA & 9659 & 7.85 & 223.88 &-4.36& 1400 & 1.3  \\
WDJ115538.39-001937.52 & 178.909362 & -0.327182 & 19.4 & DA & 5807 & 7.43 & 171.46 & -4.88& 1400 & 1.1 \\
WDJ121531.17-002255.04 & 183.879824 & -0.381983 & 19.6 & DA & 26334 & 7.68 & 744.41 & -2.99& 2050 & 4.0 \\
WDJ131038.73+324453.71 & 197.661217 & 32.748326 & 19.1 & DA & 9813 & 7.33 & 427.99 & -3.22& 2400 & 3.7 \\

\hline
\end{tabular}
\par Column (1) lists the DESI WD identifier. Columns (2) and (3) give the equatorial coordinates (R.A. and Decl.) in degrees. Column (4) presents the Gaia $G$-band apparent magnitude, and Column (5) reports the WD spectral type. Columns (6) and (7) provide the atmospheric parameters of the WD, namely the effective temperature $T_{\mathrm{eff}}$ (K) and surface gravity $\log g$ (cgs). Column (8) lists the distance in parsecs, and Column (9) gives the logarithmic luminosity relative to the Sun, $\log(L/L_{\odot})$. Columns (10) and (11) present the best-fit companion temperature derived from the two-component SED fitting, $T_{\mathrm{BT}}$ (K), and the corresponding reduced chi-square, $\chi^2_{\mathrm{BT}}$, respectively.
\\
$^a$ Reported by \cite{2011ApJS..197...38D}.
\end{table*}

\subsection{WD + BD Binary Candidates}
We identify five WD+BD binary candidates, listed in the lower part of Table~\ref{tab:two_samples}. As shown in Figure~\ref{fig:WD+BD}, these systems exhibit IR excesses that become noticeable from the NIR bands onward, consistent with the presence of a cool companion. Similar to WD+M binaries, the optical photometry of these objects is generally well reproduced by a pure WD atmosphere model, indicating that the companion contributes negligibly at optical wavelengths.

In contrast to M dwarfs, BDs have substantially lower effective temperatures, and their emission peaks at longer wavelengths. As a result, the excess in WD+BD candidates tends to rise more gradually across the NIR and often extends to longer IR wavelengths. In several cases, the IR enhancement appears as a smooth continuum increase without a pronounced turnover feature, making the overall SED shape partially degenerate with that expected from a circumstellar dust disk under the current photometric precision and wavelength coverage. Broadband SED fitting alone is therefore insufficient to unambiguously distinguish between a BD companion and a dust-disk interpretation in some systems.

To further assess the nature of these candidates, we use the Gaia low-resolution spectrophotometric data and transform them into the J-PAS narrow-band photometric system following the approach adopted in previous studies \citep{2025A&A...699A.153R}. The resulting synthetic J-PAS photometry shows no evidence for excess emission in the optical regime and remains consistent with a WD-dominated spectrum at short wavelengths. The deviation becomes apparent only toward longer wavelengths, in agreement with the expected behaviour of cool companions. We stress that this Gaia-to-J-PAS comparison serves as a qualitative consistency check and does not enter into the formal classification.

All five WD+BD candidates reported here are newly identified IR excess systems. Given the inherent degeneracies in broadband SED modelling, follow-up observations—particularly IR spectroscopy—are required to confirm the presence and properties of the proposed BD companions and to rule out alternative disk interpretations.

The occurrence rate of WD+BD candidates in our sample is $5/316 \simeq 1.6\%$. 
WD+BD systems are believed to be intrinsically rare, with reported occurrence rates typically in the range of $\sim 0.5$--$2.0\%$ in previous studies \citep{2011MNRAS.416.2768S,2011MNRAS.417.1210G}. 
The fraction derived here lies well within this range. In addition, the faint-magnitude nature of the DESI WD sample further reduces the sensitivity to very low-luminosity companions. 
The observed rate therefore likely reflects both the intrinsic scarcity of such systems and the observational challenges associated with their detection.

\subsection{WD + BD or Dust Candidates}
We identify 16 systems whose IR excess cannot be uniquely attributed to either a BD companion or a circumstellar dust disk. These objects are listed in Table~\ref{tab:BD_dust_sample}, and representative SED fits are shown in Figure~\ref{fig:BD_or_dust}. In these cases, both the WD+BD and WD+dust disk models provide comparably good fits to the available broadband photometry, resulting in very similar reduced $\chi^{2}$ values. The overall SED morphologies in the infrared are therefore insufficient to distinguish between the two physical scenarios.

The degeneracy arises because a cool substellar companion and a warm dust disk can produce broadly similar continuum enhancements over the IR wavelength range, particularly under the current photometric precision and wavelength coverage. Without distinctive spectral features or extended wavelength leverage, broadband SED fitting alone does not provide a decisive diagnostic. Discriminating between the two scenarios generally requires higher-quality infrared observations, especially spectroscopy capable of revealing molecular absorption features characteristic of BDs or constraining the long-wavelength continuum shape expected from thermal dust emission.

Given these limitations, we retain these objects as WD+BD or dust candidates without imposing a more specific classification. Their physical nature remains to be clarified by future follow-up observations.

Among the 16 systems in this category, 15 are newly identified IR-excess candidates in this work. The remaining object was previously reported by \cite{2023ApJ...944...23W}, who likewise did not reach a definitive conclusion regarding the origin of its IR excess. That study similarly noted that the available photometric data were consistent with either an L-type companion or a circumstellar dust disk. This consistency across independent analyses further highlights the intrinsic difficulty of resolving the companion--disk degeneracy using broadband photometry alone.

\begin{table*}
\centering
\scriptsize
\caption{WD+BD or Dust Candidates}
\label{tab:BD_dust_sample}
\begin{tabular}{lcccccccccccc}
\hline
 & & & & & & & & \multicolumn{2}{c}{\textbf{Companion}} & \multicolumn{2}{c}{\textbf{Blackbody}} \\
Name & R.A. & Decl. & $G$ & SpT & $T_{\rm eff}$ & $\log g$ & Dis & $T_{\rm BT}$ & $\chi^2_{\rm BT}$ & $T_{\rm BB}$ & $\chi^2_{\rm BB}$ \\
 & (deg) & (deg) & (mag) &  & (K) & (cgs) & (pc) & (K) &  & (K) &  \\
\hline
WDJ091616.71+001356.20 & 139.069421 & 0.232289 & 18.7 & DA & 12851 & 7.58 & 314.87 & 1500 & 4.9 & 1600 & 4.5 \\
WDJ101853.06-003535.70 & 154.721109 & -0.593600 & 19.3 & DC & 4977 & 7.69 & 87.41 & 1200 & 21.4 & 1650 & 20.4 \\
WDJ102203.66+824310.00 & 155.505704 & 82.718613 & 17.9 & DA & 5600 & 8.38 & 40.08 & 1400 & 4.7 & 2050 & 6.3 \\
WDJ103342.41+095027.89 & 158.426723 & 9.841012 & 19.2 & DA & 22843 & 8.85 & 261.71 & 1500 & 4.5 & 1300 & 3.3 \\
WDJ103722.63+085153.52 & 159.344220 & 8.864922 & 17.8 & DA & 17190 & 7.74 & 239.26 & 1200 & 5.5 & 1450 & 4.5 \\
WDJ114946.69-005456.32 & 177.44450 & -0.915601 & 18.39 & DA & 24520 & 7.4 & 688.75 & 1500 & 2.7 & 1350 & 2.4 \\
WDJ115816.87-012253.62 & 179.570364 & -1.381607 & 19.2 & DA & 10479 & 7.41 & 450.91 & 1400 & 0.9 & 2100 & 1.0 \\
WDJ124912.62+253959.29 & 192.302534 & 25.666382 & 19.8 & DA & 6877 & 8.13 & 164.77 & 1500 & 1.1 & 1300 & 1.0 \\
WDJ130558.49+305354.48 & 196.493435 & 30.898377 & 19.7 & DC & 7160 & 7.59 & 246.06 & 1400 & 11.6 & 2000 & 9.5 \\
WDJ133305.34+325400.11 & 203.272363 & 32.899917 & 19.0 & DBZ & 13488 & 8.04 & 296.85 & 1600 & 8.6 & 1700 & 7.0 \\
WDJ141840.40-002516.57 & 214.668362 & -0.421248 & 19.0 & DA & 15140 & 8.18 & 300.27 & 1200 & 2.5 & 1800 & 3.1 \\
WDJ142502.46-001231.96 & 216.260273 & -0.209070 & 20.2 & DA & 7153 & 8.48 & 151.32 & 1200 & 2.3 & 1400 & 2.7 \\
WDJ144028.47-012040.94 & 220.118538 & -1.344691 & 19.3 & DA & 31793 & 7.99 & 431.18 & 1400 & 4.0 & 1750 & 3.9 \\
WDJ145930.07+311609.72 & 224.875226 & 31.269303 & 19.5 & DBA & 11914 & 7.93 & 417.01 & 1500 & 5.5 & 2100 & 4.2 \\
WDJ150255.96+323942.87 & 225.732765 & 32.661963 & 19.8 & DA & 5610 & 8.16 & 116.32 & 1200 & 4.8 & 1300 & 4.5 \\
WDJ162754.97+314324.08$^a$ & 246.978929 & 31.723427 & 18.5 & DA & 18971 & 7.78 & 368.43 & 1500 & 2.4 & 2050 & 2.1 \\
\hline
\end{tabular}
\par
Columns (1)--(10) are the same as in Table~\ref{tab:two_samples}. Columns (11) and (12) give the best-fit companion temperature from the two-component SED fitting, $T_{\mathrm{eff}}(\mathrm{BT})$ (K), and the corresponding reduced chi-square, denoted as $\chi^2_{\mathrm{BT}}$, respectively. Columns (13) and (14) report the blackbody temperature of the dust-disk component, $T_{\mathrm{eff}}(\mathrm{BB})$ (K), and the corresponding reduced chi-square, denoted as $\chi^2_{\mathrm{BB}}$, respectively.\\
$^a$ Reported by \cite{2023ApJ...944...23W}.
\end{table*}

\subsection{WD + Dust Disk Candidates}
\begin{figure}
    \centering
    \includegraphics[width=0.75\linewidth]{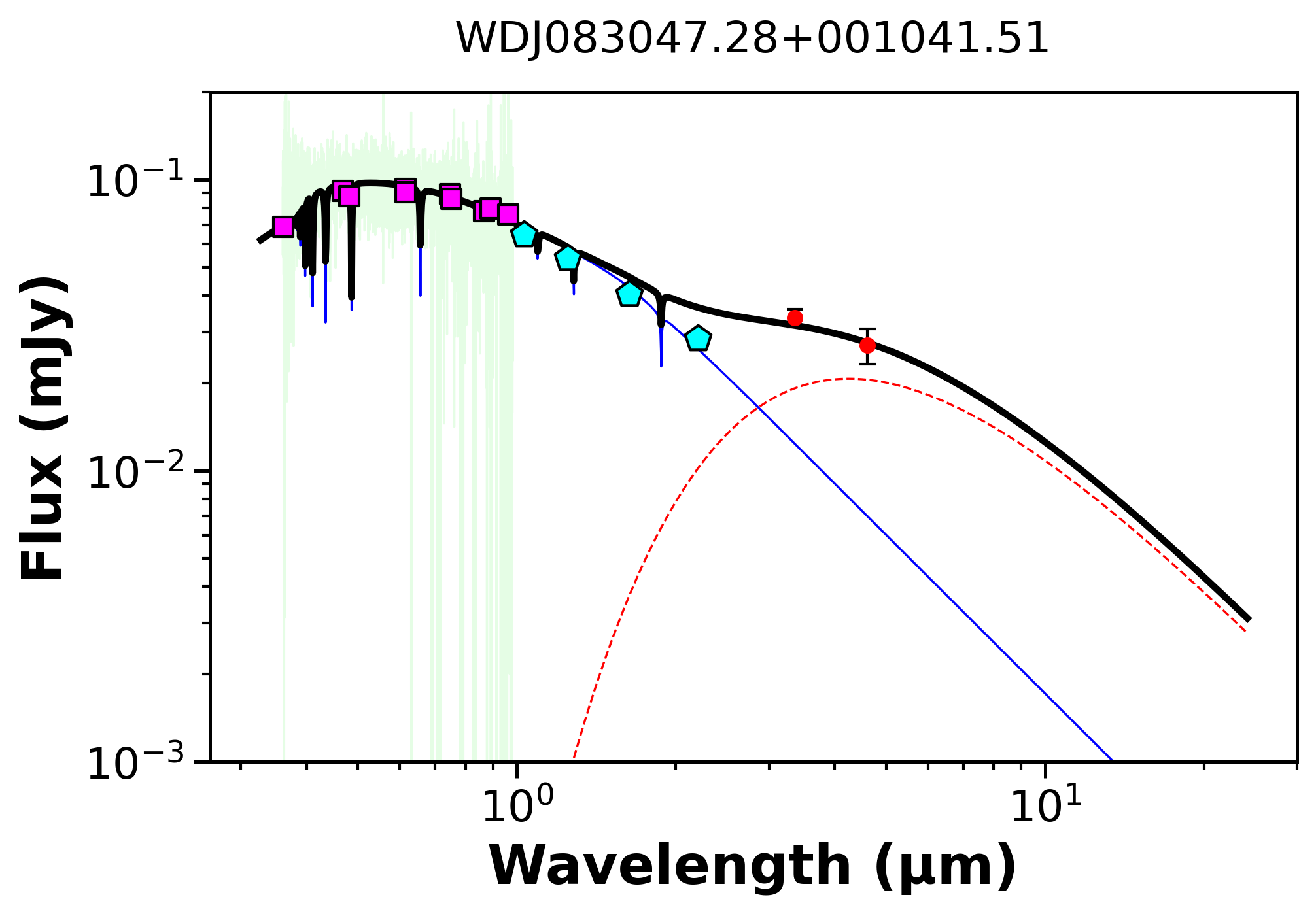}
    \caption{SED fitting result for the WD+dust disk candidate WDJ083047.28+001041.51. The best-fitting model (black line) consists of a WD atmosphere model (Koester; blue line) combined with a circumstellar dust component represented by a single-temperature blackbody (red dashed line), yielding reduced $\chi^2 = 2.9$. Photometric measurements are taken from SDSS or Pan-STARRS (magenta squares), UKIDSS or 2MASS (cyan pentagons), and \textit{WISE} (red circles). The light-green curve shows the DESI optical spectrum, which is well reproduced by the WD model. The excess emission becomes apparent at IR wavelengths and is consistent with thermal emission from a dust disk.
}
    \label{fig:WD+dust}
\end{figure}

We identify 38 WD+dust disk candidates, whose basic properties and fitting parameters are summarized in Table~\ref{tab:dust_sample}. As illustrated in Figure~\ref{fig:WD+dust}, the IR excess in these systems typically becomes prominent at the IR wavelengths, particularly in the \textit{WISE} W1/W2 bands. Such behaviour is consistent with thermal emission from dust at characteristic temperatures of order $\sim 1000$\,K. In contrast to companion-dominated systems, little or no significant excess is usually present in the NIR, supporting an interpretation in which the excess arises from circumstellar dust rather than from a hotter secondary component.

Among the 38 candidates, 28 are newly identified IR excess systems in this work. The remaining 10 objects have previously been discussed in the literature. Three have been confirmed as debris disk systems based on \textit{Spitzer} observations; two were reported as WD+dust disk candidates; and several others were noted to exhibit IR excess without a definitive interpretation. In one case, a possible L-type companion was suggested, although no firm conclusion was reached. 

One source was previously flagged as potentially affected by source confusion. \cite{2020ApJ...891...97D} noted that its \textit{Spitzer}/IRAC Ch~1 and Ch~2 fluxes were ``confused with a background galaxy,'' and the corresponding \textit{WISE} excess was considered uncertain. We independently inspected the available optical and NIR images and did not identify obvious nearby contaminants. The \textit{WISE} data-quality flags (e.g., $cc\_flags=0000$) likewise show no clear indications of artifacts. Nevertheless, given the relatively large \textit{WISE} point-spread function, contamination from a faint, very red background source cannot be fully excluded. We therefore retain this object in our sample while acknowledging that its IR excess nature requires further confirmation with higher-resolution IR imaging or spectroscopy.

\begin{figure}
    \centering
    \includegraphics[width=0.75\linewidth]{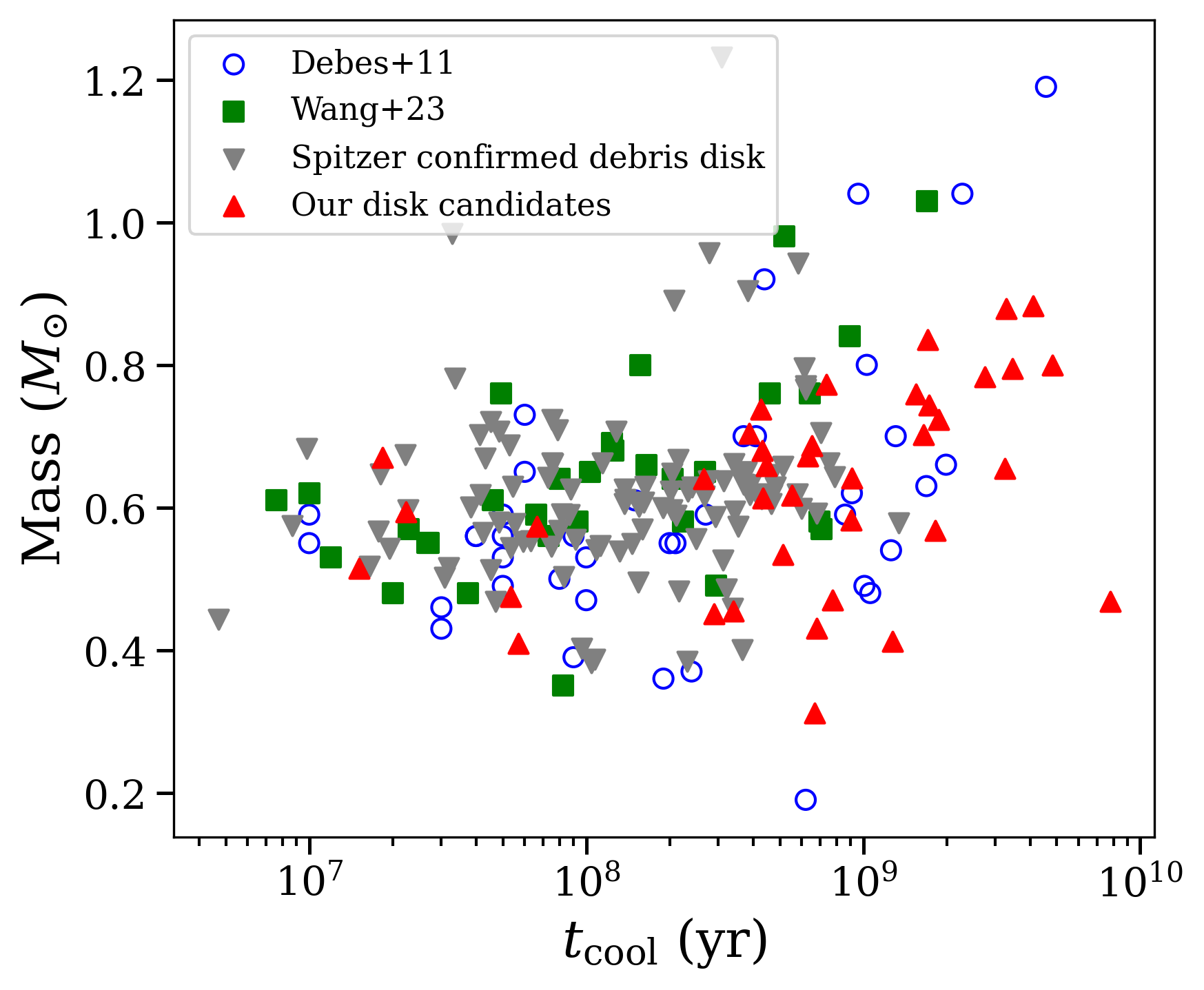}
    \caption{Mass--cooling age diagram for WDs with IR excess attributed to circumstellar dust disks. Red triangles mark the 38 disk candidates identified in this work, while literature samples are shown for reference: blue open circles \citep{2011ApJS..197...38D} , green squares \citep{2023ApJ...944...23W}, and gray inverted triangles (disk systems confirmed by \textit{Spitzer}). Compared to previous compilations, our candidates populate similar mass ranges but preferentially extend to longer cooling ages, thereby broadening the parameter space of known dusty disk WDs.}
    \label{fig:MASS_Vs_age}
\end{figure}

To place these systems in an evolutionary context, we examine the distributions of WD mass and cooling age and compare them with previously reported dusty disk samples, including those from \cite{2011ApJS..197...38D}, \cite{2023ApJ...944...23W}, and debris-disk systems confirmed by \textit{Spitzer} (Figure~\ref{fig:MASS_Vs_age}). Our candidates extend toward relatively older cooling ages compared to several earlier samples. This difference may reflect the faint-magnitude focus and selection characteristics of the DESI WD sample, although it also leaves open the possibility that dusty disks can persist or be replenished at later evolutionary stages.

The mass distribution of our sample is broadly consistent with those reported in previous studies, while the cooling-age range extends to older systems.

The occurrence rate of WD+dust disk candidates in our sample is $38/316 \simeq 12\%$, higher than the typical debris-disk fractions of $\sim 1$--6\% reported in earlier work \citep{2015MNRAS.449..574R,2020ApJ...902..127X,2024A&A...688A.168M,Wang_2026}. This value should be interpreted as the candidate fraction under our current selection criteria rather than as a direct measurement of the intrinsic disk incidence. The faintness of the parent sample and the modest angular resolution of \textit{WISE} both increase the susceptibility to small photometric offsets and source blending, which can enhance the apparent excess frequency. Although imaging inspections and contamination checks have been applied, some borderline cases may remain. A robust determination of the intrinsic debris-disk fraction will require confirmation through higher-resolution imaging and IR spectroscopy.

\begin{table*}
\centering
\scriptsize  
\caption{WD + Dust Disk Candidates}
\label{tab:dust_sample}
\begin{tabular}{lccccccccccc}
\hline
Name & R.A. & Decl. & $G$ & SpT & $T_{\rm eff}$ & $\log g$ & Mass & Age & Dis & $T_{BB}$ & $\chi^2_{BB}$ \\
(DESI) & (deg) & (deg) & (mag) &  & (K) & (cgs) & ($M_\odot$) & (Gyr) & (pc) & (K) &  \\
\hline
WDJ001249.08+094259.81 & 3.204681 & 9.716345 & 20.0 & DC & 6729 & 8.32 & 0.79 & 3.46 & 157.62 & 1050 & 7.1 \\
WDJ001321.07+282019.83$^a$ & 3.338118 & 28.338805 & 15.7 & DA & 25155 & 7.76 & 0.52 & 0.43 & 133.91 & 1400 & 10.2 \\
WDJ022320.55-045906.66$^c$ & 35.835757 & -4.985243 & 16.8 & DA & 10632 & 8.14 & 0.69 & 0.43 & 83.46 & 500 & 17.8 \\
WDJ083047.28+001041.51$^b$ & 127.696934 & 0.178144 & 18.9 & DA & 8230 & 7.76 & 0.47 & 0.02 & 204.13 & 1200 & 2.4 \\
WDJ083233.92+814938.22$^a$ & 128.140576 & 81.827175 & 16.3 & DA & 23851 & 7.93 & 0.59 & 0.05 & 150.18 & 800 & 1.6 \\
WDJ084949.62+092353.46$^{e,f}$ & 132.456817 & 9.398091 & 17.6 & DA & 8569 & 7.98 & 0.58 & 0.65 & 98.93 & 450 & 4.9 \\
WDJ085410.82+091835.00 & 133.545250 & 9.309408 & 19.8 & DA & 7447 & 8.21 & 0.72 & 0.78 & 186.95 & 1000 & 0.8 \\
WDJ091446.69+002331.65 & 138.694480 & 0.392070 & 18.7 & DA & 16582 & 7.49 & 0.41 & 0.02 & 368.03 & 750 & 1.4 \\
WDJ092031.23+330831.41 & 140.130230 & 33.142026 & 19.9 & DC & 8283 & 7.65 & 0.43 & 0.91 & 333.09 & 1150 & 2.4 \\
WDJ092505.44+314836.75 & 141.272651 & 31.810115 & 18.9 & DA & 10948 & 7.70 & 0.45 & 1.87 & 251.84 & 950 & 2.1 \\
WDJ100657.72+015203.60 & 151.740360 & 1.867446 & 19.4 & DC & 7596 & 8.18 & 0.70 & 0.06 & 159.39 & 1000 & 1.1 \\
WDJ120558.25-004217.11 & 181.492981 & -0.705174 & 18.7 & DC & 7150 & 8.30 & 0.78 & 0.68 & 102.19 & 850 & 3.5 \\
WDJ121209.31+013627.77$^d$ & 183.038676 & 1.607441 & 18.0 & DAH & 10668 & 8.03 & 0.62 & 0.34 & 149.69 & 1150 & 4.2 \\
WDJ121653.30+745525.29$^c$ & 184.221715 & 74.924041 & 16.8 & DAZ & 6467 & 7.96 & 0.57 & 0.45 & 41.45 & 1100 & 4.1 \\
WDJ124804.03+282104.11 & 192.016830 & 28.350956 & 18.0 & DA & 13101 & 8.16 & 0.70 & 1.65 & 183.16 & 1150 & 5.4 \\
WDJ125331.45+265806.70 & 193.381120 & 26.968484 & 19.2 & DA & 7443 & 8.44 & 0.88 & 2.74 & 118.12 & 1200 & 3.1 \\
WDJ125414.48+274533.35 & 193.560201 & 27.759214 & 18.8 & DA & 10235 & 7.88 & 0.53 & 0.55 & 228.77 & 1100 & 4.8 \\
WDJ125520.03+272524.32 & 193.833446 & 27.423445 & 18.0 & DA & 11633 & 8.01 & 0.61 & 1.82 & 165.16 & 1000 & 3.7 \\
WDJ125847.33+233844.37$^e$ & 194.696912 & 23.645520 & 19.0 & DA & 7820 & 8.24 & 0.74 & 0.39 & 129.75 & 1200 & 15.0 \\
WDJ125854.90+320428.51 & 194.728811 & 32.074713 & 19.1 & DA & 6303 & 7.63 & 0.41 & 3.28 & 143.68 & 550 & 2.8 \\
WDJ125931.57+263907.62 & 194.881420 & 26.652077 & 19.8 & DA & 9012 & 8.07 & 0.64 & 0.51 & 231.73 & 800 & 1.4 \\
WDJ130303.73+233418.47 & 195.765599 & 23.571747 & 19.7 & DA & 11468 & 7.69 & 0.45 & 0.43 & 474.97 & 1100 & 1.0 \\
WDJ130401.13+255546.14 & 196.004634 & 25.929524 & 19.3 & DC & 8703 & 8.38 & 0.84 & 1.73 & 162.54 & 1150 & 3.4 \\
WDJ133140.57+310032.03 & 202.918325 & 31.009019 & 19.7 & DAH & 5987 & 8.33 & 0.80 & 1.28 & 109.84 & 1050 & 8.7 \\
WDJ135828.55+054632.38 & 209.618998 & 5.775717 & 19.3 & DA & 6825 & 8.45 & 0.88 & 0.91 & 103.31 & 450 & 3.6 \\
WDJ135833.07+051501.99 & 209.637774 & 5.250604 & 19.1 & DA & 10968 & 8.27 & 0.77 & 0.29 & 244.90 & 800 & 2.1 \\
WDJ140006.82+054719.29 & 210.028200 & 5.788454 & 18.2 & DA & 7259 & 7.33 & 0.31 & 1.71 & 148.96 & 850 & 4.1 \\
WDJ141943.13-000836.30 & 214.929998 & -0.144381 & 18.7 & DA & 5871 & 8.11 & 0.66 & 4.82 & 74.73 & 1100 & 1.2 \\
WDJ143333.09+030819.09 & 218.387827 & 3.138614 & 19.5 & DA & 8225 & 8.26 & 0.76 & 4.12 & 172.03 & 900 & 2.5 \\
WDJ143406.77+150817.81$^{e,f}$ & 218.528051 & 15.138229 & 16.0 & DA & 14129 & 8.05 & 0.64 & 0.63 & 79.91 & 400 & 26.6 \\
WDJ143613.04+011650.67 & 219.054507 & 1.280252 & 19.1 & DA & 6174 & 7.79 & 0.47 & 0.74 & 112.44 & 1150 & 3.2 \\
WDJ143751.82+005254.99 & 219.465776 & 0.882108 & 17.7 & DA & 19033 & 7.91 & 0.57 & 0.67 & 240.87 & 500 & 4.3 \\
WDJ144334.64-015340.39 & 220.894343 & -1.894512 & 19.4 & DA & 13014 & 8.21 & 0.74 & 3.24 & 304.32 & 550 & 1.7 \\
WDJ144433.81-005958.87 & 221.141003 & -0.999585 & 16.3 & DA & 12333 & 8.12 & 0.68 & 1.55 & 77.79 & 1050 & 13.7 \\
WDJ150139.44+312245.83 & 225.414206 & 31.379377 & 19.5 & DA & 18103 & 7.69 & 0.47 & 0.27 & 579.42 & 1200 & 0.8 \\
WDJ161316.63+552125.97$^a$ & 243.319090 & 55.357180 & 15.9 & DAZ & 11980 & 8.09 & 0.66 & 0.02 & 66.42 & 450 & 40.4 \\
WDJ163237.16+310518.97 & 248.154712 & 31.088682 & 18.8 & DAZ & 10645 & 8.12 & 0.67 & 1.60 & 215.14 & 450 & 4.6 \\
WDJ164407.41+303143.25 & 251.030822 & 30.528611 & 17.9 & DA & 26057 & 8.06 & 0.67 & 0.07 & 314.18 & 1100 & 4.7 \\
\hline
\end{tabular}
\par
Columns (1)--(8) are the same as in Table~\ref{tab:two_samples}. Columns (9) and (10) give the dust-disk blackbody temperature, $T_{\mathrm{eff}}(\mathrm{BB})$ (K), and the corresponding reduced chi-square, denoted as $\chi^2_{\mathrm{dust}}$, respectively.\\
$^a$ are previously confirmed debris disk systems identified by \textit{Spitzer}.\\
$^b$ Reported by\cite{2020ApJ...891...97D}.\\
$^c$ Reported by \cite{2019MNRAS.489.3990R}. \\
$^d$ Reported by \cite{2011MNRAS.417.1210G}.\\
$^e$ Reported by \cite{2011ApJS..197...38D}.\\
 $^f$ reported as a candidate gaseous debris disk\citep{2025AJ....170..345M}.
\end{table*}


\section{Conclusions}
\label{sect:conclusion}
We performed a systematic search for IR excess among 2706 spectroscopically confirmed WDs from the DESI Early Data Release. By combining optical, IR photometry and fitting SEDs with conservative excess criteria, we assembled a homogeneous set of IR-excess WD candidates. We then evaluated the robustness of the excess emission using imaging-based contamination checks and two-component SED models to probe its likely physical origin.

Starting from 316 WDs with complete multi-wavelength coverage, we identified 72 preliminary IR-excess candidates. After removing sources compromised by blending or confusion, 62 systems remain as our final working sample. Two-component SED fitting places these objects into four groups: three WD+M dwarf binary candidates, five WD+BD binary candidates, 38 WD+dust disk candidates, and 16 systems whose excess cannot be uniquely attributed to either a cool companion or a circumstellar disk with the current photometric constraints.

These candidate WD+M and WD+BD systems are relevant for testing the survival of low-mass companions through post-main-sequence evolution and for constraining outcomes of CE evolution. The disk-bearing candidates extend the parameter space of dusty WDs toward fainter magnitudes and older cooling ages, offering a view of the long-term dynamical evolution of planetary remnants around WDs. The ambiguous subset, meanwhile, underscores a fundamental limitation of broadband SED fitting: companion and disk scenarios can produce similar colors when the excess is weak or emerges primarily in the IR.

Breaking this degeneracy requires targeted follow-up. IR spectroscopy can directly reveal molecular absorption features expected from BDs or constrain the continuum shape characteristic of thermal dust emission, while higher-resolution imaging can mitigate source confusion and identify unresolved contaminants. Such observations will be essential both for confirming the nature of individual systems and for refining intrinsic occurrence rates of low-mass companions and debris disks in the DESI WD population.

\begin{acknowledgements}
We thank the anonymous referee for helpful suggestions that improved the quality of the paper. We thank Siyi Xu for helpful discussions and valuable suggestions on this work. This work was supported in part by the National Natural Science Foundation of China under grant U1631109. This work was also supported in part by the Guizhou Provincial Basic Research Program (Natural Science) (No. MS[2025]694), the Guizhou Provincial Major Scientific and Technological Programs (No. XKBF (2025)011, XKBF (2025)010) and the Guizhou Provincial Science and Technology Projects (No. QKHFQ[2023]003 and QKHPTRC-ZDSYS[2023]003). 
\end{acknowledgements}

\label{lastpage}
\bibliographystyle{raa}
\bibliography{ref}

\end{document}